\documentclass[twocolumn,pre,epsfig,rotate,noshowpacs,cjk,english]{revtex4}
\usepackage{graphicx}
\usepackage{amsmath}
\usepackage{dcolumn}
\usepackage{epsfig}
\usepackage{bm}
\begin{document}

\title{Heat perturbations spreading in the Fermi-Pasta-Ulam-$\beta$ system with next-nearest-neighbor coupling: Competition between phonon dispersion and nonlinearity}

\author{Daxing Xiong}
\email{phyxiongdx@fzu.edu.cn}
\affiliation{Department of Physics,
Fuzhou University, Fuzhou 350108, Fujian, China}

\begin{abstract}
We employ the heat perturbations correlation function to study
thermal transport in the one-dimensional (1D)
Fermi-Pasta-Ulam-$\beta$ lattice with both nearest-neighbor and
next-nearest-neighbor couplings. We find that such a system bears a
peculiar phonon dispersion relation, and thus there exists a
competition between phonon dispersion and nonlinearity that can
strongly affect the heat correlation function' shape and scaling
property. Specifically, for small and large anharmoncity, the
scaling laws are ballistic and superdiffusive types, respectively,
which are in good agreement with the recent theoretical predictions;
whereas in the intermediate range of the nonlinearity, we observe an
unusual multiscaling property characterized by a \emph{nonmonotonic}
delocalization process of the central peak of the heat correlation
function. To understand this multiscaling laws, we also examine the
momentum perturbations correlation function and find a transition
process with the same turning point of the anharmonicity as that
shown in the heat correlation function. This suggests coupling
between the momentum transport and the heat transport, in agreement
with the theoretical arguments of mode cascade theory.
\end{abstract}
\maketitle

\section{Introduction}
Despite decades of research, our understanding of anomalous heat
transport in one-dimensional (1D) momentum-conserving systems is
still scarce~\cite{Lepri_Report, Dhar_Report,Lepri_Book}. There are
various theoretical models and three main numerical approaches which
devote to solving this issue. The theoretical techniques include the
mode coupling theory~\cite{ModeCoupling-1,ModeCoupling-2}, the
renormalization group method~\cite{Renoramlized}, the mode cascade
assumption~\cite{Daswell-1,Daswell-2,Daswell-3,Daswell-4}, the
phenomenological L\'evy walks
model~\cite{LevyWalks-Review,HeatPerturbations-3}, and the nonlinear
fluctuating hydrodynamics
theory~\cite{HeatPerturbations-5,HeatPerturbations-6}, etc. The
early two numerical approaches are based on either the direct
nonequilibrium molecular dynamics
simulations~\cite{MethodOne-1,MethodOne-2,MethodOne-3,MethodOne-4,MethodOne-5}
or the Green-Kubo
formula~\cite{MethodTwo-1,MethodTwo-2,MethodTwo-3}. In the former,
the ends of the system are first connected to heat baths with a
small temperature difference for a long time. After a stead state
has been obtained, one then observes the heat current flowing across
the system, and finally derives the thermal conductivity. In the
latter, one usually examines the long time asymptotic behavior of
the heat current autocorrelation function (or power spectra), and
then use Green-Kubo formula to get the heat conductivity. In this
respect, the first three theoretical
models~\cite{ModeCoupling-1,ModeCoupling-2,Renoramlized,Daswell-1,Daswell-2,Daswell-3,Daswell-4}
are just devoted to prediction of the time scaling behavior of the
heat current autocorrelation function. In particular, the mode
cascade theory suggested for the first time that one should take the
coupling between the momentum transport and the heat transport into
account, and demonstrated that due to this coupling, to obtain an
accurate prediction, the heat current power spectra at sufficiently
low frequencies should be
probed~\cite{Daswell-1,Daswell-2,Daswell-3,Daswell-4}.

The third numerical approach is based on the perturbations
correlation functions. This is inspired by the idea of diffusion of
energy in the lattice systems. As is well known, in such many-body
systems, particles are mainly located around their equilibrium
positions. So there is no sense to talk about the diffusion of
energy of the associated particles. While the collective system
dynamics creates a ``tissue'', which can react to small local
perturbations affecting its dynamics. Eventually, the propagation of
perturbations defines the overall information of
transport~\cite{LevyWalks-Review,Perturbations-1,Perturbations-2,Perturbations-3,Perturbations-4}.

With such an idea in mind, understanding the energy and heat
transport processes via their perturbations correlation functions is
a relevant fascinating topic of theoretical research~\cite{
HeatPerturbations-1,HeatPerturbations-2,LevyWalks-Review,HeatPerturbations-3,HeatPerturbations-4,HeatPerturbations-5,HeatPerturbations-6,HeatPerturbations-7,HeatPerturbations-8,
HeatPerturbations-9,HeatPerturbations-10,HeatPerturbations-11,Xiong-1,Xiong-2,Xiong-3}.
The early numerical works indicated that for general 1D nonlinear
nonintegrable momentum-conserving systems, a quasi superdiffusive
L\'evy walks profile of the energy perturbations correlation
function can be always observed
~\cite{HeatPerturbations-1,HeatPerturbations-2,HeatPerturbations-3,
HeatPerturbations-4}, which is the evidence of anomalous heat
transport. Viewing this fact, such superdiffusive transport has been
subsequently understood from the single particle's L\'evy walks
model in the superdiffusive regime after considering the particle's
velocity fluctuations~\cite{LevyWalks-Review,HeatPerturbations-3},
although the connection between them is only
phenomenological~\cite{LevyWalks-Review}. Later, a more detailed
mechanism has been considered in a broad context of
hydrodynamics~\cite{HeatPerturbations-5, HeatPerturbations-6}, where
the authors developed a nonlinear version of the hydrodynamics
theory and explained the observed L\'evy walks profile as a
combination of the heat and sound modes'
correlations~\cite{HeatPerturbations-4,
HeatPerturbations-5,HeatPerturbations-6}. The main achievement of
this hydrodynamics theory is that it can be used to predict the
scaling property of the heat and momentum perturbations spreading
correlation functions in certain nonlinear systems and thus is
greatly helpful to our understanding of anomalous heat
transport~\cite{HeatPerturbations-7}. This is because the heat and
the sound modes' correlations are usually conjectured to correspond
to the heat and the momentum perturbations correlation functions,
respectively~\cite{HeatPerturbations-4}. In particular, for generic
systems of nonzero pressure, such as the
Fermi-Pasta-Ulam-$\alpha$-$\beta$ (FPU-$\alpha$-$\beta$) model with
asymmetric (odd) interparticle potential, the prediction for the
sound modes' correlation is Kardar-Parisi-Zhang (KPZ)
scaling~\cite{KPZ}; while for the particular FPU-$\beta$ system with
symmetric (even) interparticle potential, it is not KPZ but Gaussian
scaling~\cite{HeatPerturbations-4, HeatPerturbations-5}. This is
consistent with the argument that the momentum current power
spectrum does (does not) diverge at low frequency for systems with
odd (even) interparticle
potential~\cite{ModeCoupling-1,Daswell-1,Daswell-2,Daswell-3,Daswell-4}.

In spite of these achievements, an unclear point from the nonlinear
hydrodynamics theory may be that, we still don't know to what extent
of the strength of the nonlinearity it would be applicable. Since
the theory aims to deal with the nonlinear systems, it requires the
system's dynamics to be sufficiently chaotic so as to have good
mixing in time~\cite{HeatPerturbations-6}. So it is quite possible
that, for certain linear integrable systems, the nonlinear
hydrodynamics theory is not valid any
more~\cite{HeatPerturbations-11}. As to the latter systems, a recent
concept called ``phonon random walks'' might be
worthwhile~\cite{PRW}, based on which the linear system's ballistic
heat perturbations spreading correlation function can be instead
understood by a quantum-like wave function's modulus square, where
the different systems' distinct phonon dispersion relations are a
key factor. In addition, in this theory the momentum perturbations
correlation function is proved to be the corresponding wave
function's real part~\cite{PRW}.

Combining the above reviewing progress, one may recognize that the
coupling between the heat transport and the momentum transport
should naturally exist and this coupling will be exhibited quite
differently in various systems. This then leads to the fact that
different theoretical models would be applicable to different
systems or to the same system under different strength of the
nonlinearity. In particular, for the strongly nonlinear systems
where the effects of nonlinearity dominate, the nonlinear
hydrodynamics theory~\cite{HeatPerturbations-5, HeatPerturbations-6}
could present some universal predictions; while as to the linear
systems without including the nonlinearity, the phonon dispersion
relation now plays a major role~\cite{PRW} and there is not mode and
mode's coupling~\cite{Linear}. Therefore, to provide a complete
picture of thermal transport, such as the normal, ballistic,
superdiffusive types, and to understand the coupling between the
heat transport and the momentum transport in different situations,
both the effects of nonlinearity and phonon dispersion relation are
crucial and necessary to take into account. Motivated by this and in
view of that almost all of above literatures only focused on the
systems with nearest-neighbor (NN) interaction, we here consider a
1D FPU-$\beta$ system with both the NN and next-nearest-neighbor
(NNN) interactions. The advantage of this system is that it bears a
peculiar phonon dispersion relation. With this advantage, we then
can adjust the strength of the nonlinearity to see how such
particular phonon dispersion relation could together with the
nonlinearity affect the system's heat transport and its scaling
property. Such a research strategy will also help us to reveal the
detailed coupling between the heat transport and the momentum
transport in this particular system, which may provide insights into
further developing a theory to bridge our present understanding of
linear and nonlinear systems' thermal transport, after combining
both factors of phonon dispersion relation and nonlinearity.

The rest of this paper is structured in the following way: In Sec.
II we introduce the reference model and emphasize its peculiar
phonon dispersion relation. Sec. III describes the simulation method
to derive the corresponding perturbations correlation functions. In
Sec. IV we present our main results, and Sec. V is devoted to
discussion of the mechanism. Finally we close with a summary in Sec.
VI.
\section{Model}
%%%%%%%%%%%%%%%%%%%%%%%%%%%%%%%%%%%%%%%%%%%%%%%%%%%%%%%%%%%%%%%%%%%%%%%%%%%%%%%%fig1
\begin{figure}
\vskip-.2cm \hskip-0.4cm
\includegraphics[width=8.8cm]{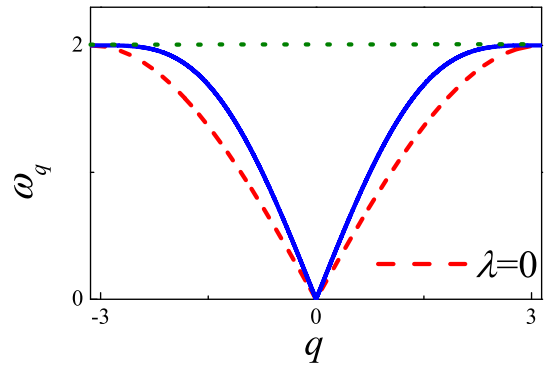}
\vskip-0.4cm \caption{\label{Fig1} The phonon dispersion relation
for the FPU-$\beta$ chain with both NNN and NN couplings
($\lambda=0.25$), which is compared to the counterpart harmonic
chain of $\lambda=0$ (dashed line). The horizontal dotted line
indicates $\omega_q=2$ near the Brillouin zone boundary.}
\end{figure}
%%%%%%%%%%%%%%%%%%%%%%%%%%%%%%%%%%%%%%%%%%%%%%%%%%%%%%%%%%%%%%%%%%%%%%%%%%%%%%%%fig1
As mentioned, in what follows we focus on a 1D FPU-$\beta$ lattice
with both NN and NNN interactions~\cite{NNN}. Such a system's
Hamiltonian is
\begin{equation} \label{Hamiltonian}
H= \sum_{k=1}^{L} \left[ \frac{p_{k}^2}{2}+ V(r_{k+1}-r_k) +\lambda
V(r_{k+2}-r_k) \right]
\end{equation}
with $r_k$ the displacement of the $k$th particle from its equilibrium position and $p_k$ its momentum. The potential takes the FPU-$\beta$ type of $V(\xi)=\xi^2/2+\beta \xi^4/4$ with $\beta$ to be adjustable and representing the nonlinearity. The parameter $\lambda$ controls the comparative strength of the NNN coupling to the NN coupling and if it is fixed at $\lambda =0.25$ reproducing a special phonon dispersion relation (under harmonic approximation):
\begin{equation} \label{dispersion}
\omega_{q} = \sqrt{4 \sin^2\left(q/2\right)+ \sin^2\left( q \right)}
\end{equation}
as shown in Fig.~\ref{Fig1}. Here $q$ is the wave number and
$\omega_q$ is the corresponding frequency. From Fig.~\ref{Fig1} we
know that such a phonon dispersion relation has a main feature,
i.e., the group velocity $v_g = d \omega_q/d q$ is very close to
zero in a wider $q$ domain near the Brillouin zone boundary. This
unusual property can favor the formation of a special highly
localized excitation [intraband discrete breathers (DBs)]~\cite{DBs}
in the presence of appropriate nonlinearity and thus is conjectured
to greatly influence thermal transport~\cite{NNN}. In addition, we
note that some recent works have indicated that the FPU-$\beta$
lattice including such kind of long range interactions beyond the NN
couplings can lead to some unusual effects on thermal
transport~\cite{NNN-1} and thermal rectification~\cite{NNN-2}. All
of these understanding are also the motivations that we choose to
study such a system.

\section{Method}
We shall employ the equilibrium fluctuation-correlation
method~\cite{HeatPerturbations-1, HeatPerturbations-4} to
investigate the propagation of heat perturbations. This approach has
been first proposed by Zhao~\cite{HeatPerturbations-1} for studying
the site-site total energy fluctuations spreading and then extended
to be applicable to investigate the space-space fluctuations
spreading~\cite{HeatPerturbations-4}. For further detailed
implementation, one can also refer to~\cite{PingHuang}. Due to the
apparent advantage of avoiding the emerging statistical
fluctuations~\cite{LevyWalks-Review,HeatPerturbations-3}, such
popular efficient simulation method has been widely used in many
publications~\cite{HeatPerturbations-2, HeatPerturbations-3,
HeatPerturbations-8, HeatPerturbations-9,
HeatPerturbations-10,Xiong-1,Xiong-2,Xiong-3}.

To make a comparison with the prediction of hydrodynamics theory~\cite{HeatPerturbations-5, HeatPerturbations-6}, we shall focus on the following two space-space correlation functions of the heat perturbations and momentum perturbations~\cite{HeatPerturbations-4}, i.e.,
\begin{equation}
\rho_{Q} (m,t)=\frac{\langle \Delta Q_{j}(t) \Delta Q_{i}(0)
\rangle}{\langle \Delta Q_{i}(0) \Delta Q_{i}(0) \rangle}
\end{equation}
and
\begin{equation}
\rho_{p}(m,t)=\frac{\langle \Delta p_{j}(t) \Delta p_{i}(0)
\rangle}{\langle \Delta p_{i}(0) \Delta p_{i}(0) \rangle},
\end{equation}
respectively. Here $m=j-i$; $\langle \cdot \rangle$ denotes the
spatiotemporal average. To describe the space-space correlation, one
can divide the 1D lattice into several equivalent bins and thus, $i$
and $j$ are the labels of the bins. In practice, we will set the
averaged number of particles in each bin to be $N_i=(L-1)/b$ with
$b$ the total number of the bins. Under such space description,
$\Delta Q_{i}(t)\equiv Q_i(t)- \langle Q_i \rangle$ and $ \Delta
p_{i}(t) \equiv p_i(t)- \langle p_i \rangle$ then define the heat
perturbations and momentum perturbations in the $i$th bin at time
$t$, respectively, with $Q_i(t)$ and $p_i(t)$ the corresponding heat
energy and momentum densities. $p_{i}(t)$ is readily to calculate,
one just needs to sum the related single particle's momentum
$p_k(t)$ within the bin, namely $p_{i}(t) \equiv \sum_k p_k(t)$. To
compute $Q_i(t)$, we employ the definition of $Q_i(t) \equiv
E_i(t)-\frac{(\langle E \rangle +\langle F \rangle) M_i(t)}{\langle
M \rangle}$~\cite{Forster,Liquid} from thermodynamics, where
$E_{i}(t) \equiv \sum_k E_k(t)$, $M_i(t) \equiv \sum_k M_k(t)$, and
$F_i(t) \equiv \sum_k F_k(t)$ are the  total energy, number of
particles and internal  pressure in that bin, respectively, with
$E_k(t)$, $M_k(t)$, $F_k(t)$ the corresponding single particle's
energy, density, and pressure. From the perspective of hydrodynamics
theory, $\rho_{Q} (m,t)$ and $\rho_{p} (m,t)$ might represent the
heat mode's and sound modes' correlations,
respectively~\cite{HeatPerturbations-4,HeatPerturbations-5,HeatPerturbations-6},
based on which one might be able to construct the corresponding
energy and particle perturbations correlation
functions~\cite{HeatPerturbations-4}. Therefore, our following study
of $\rho_{Q} (m,t)$ and $\rho_{p} (m,t)$ is also partially motivated
by this connection.

The simulations of both correlation functions are performed as
follows: Initially we contact the system with a Langevin heat
bath~\cite{Lepri_Report, Dhar_Report} of temperature $T=0.5$ (fixed
throughout the paper) to get a equilibrium state. Then after this
thermalized equilibrium state has been prepared, we utilize the
Runge-Kutta integration algorithm of seventh to eighth order with a
time step $h=0.05$ to evolve the system. During such evolution, we
then sample the relevant data and calculate the corresponding
correlation functions.

To perform the simulations, we apply the following settings:
periodic boundary conditions with a size of $L=4001-6001$ is
adopted, which will allow a perturbation of heat and momentum
located at the center of the chain to spread out a long time up to
$t=900$ for different $\beta$. $\beta$ is adjusted in a wide range
from $\beta=0$ (linear case) to $\beta=1.5$ (representing the highly
nonlinear case). The number of bin is fixed at $b \equiv (L-1)/2$
and has been verified to be efficient to derive the space-space
correlation information. The size of the ensemble for detecting both
correlation functions is about $8 \times 10^9$.
\section{Results}
\subsection{Scaling for linear and highly nonlinear cases}
We start with studying the following two limiting cases. The first one is the linear system with $\beta=0$, for which one can use the formula from the theory of phonon random walks~\cite{PRW}
\begin{equation} \label{PRWdensity}
\rho_Q (m,t) \simeq \left|\frac{1}{2 \pi} \int_{-\pi}^{\pi} e^{\rm{i} \left(\mit{m} \mit{q}- \omega_{q}
\mit{t} \right)} \rm{d} \mit{q}\right|^{2}
\end{equation}
to predict $\rho_Q(m,t)$. Inserting the phonon
dispersion~\eqref{dispersion} into formula~\eqref{PRWdensity}, one
then gains the prediction of $\rho_Q(m,t)$, which has been well
verified by simulations~\cite{PRW}. The second is the highly
nonlinear case. For such system, the nonlinear hydrodynamics
theory~\cite{HeatPerturbations-5,HeatPerturbations-6} has predicted
the heat mode's correlation function to be the L\'evy distribution
and satisfying the following scaling
property~\cite{LevyWalks-Review,HeatPerturbations-3}
\begin{equation}
\label{scaling} t^{1/\gamma} \rho_{Q} (m,t) \simeq \rho_{Q}
(\frac{m}{t^{1/\gamma}},t)
\end{equation}
with $\gamma=3/2$ the predicted scaling exponent. Note that such a prediction of $\gamma$ value requires the associated potential of the system to be symmetric and the internal averaged pressure $\langle F \rangle$ to be zero, our focused system here naturally bears such property.
%%%%%%%%%%%%%%%%%%%%%%%%%%%%%%%%%%%%%%%%%%%%%%%%%%%%%%%%%%%%%%%%%%%%%%%%%%%%%%%%fig2
\begin{figure}
\vskip-.2cm \hskip-0.4cm
\includegraphics[width=8.8cm]{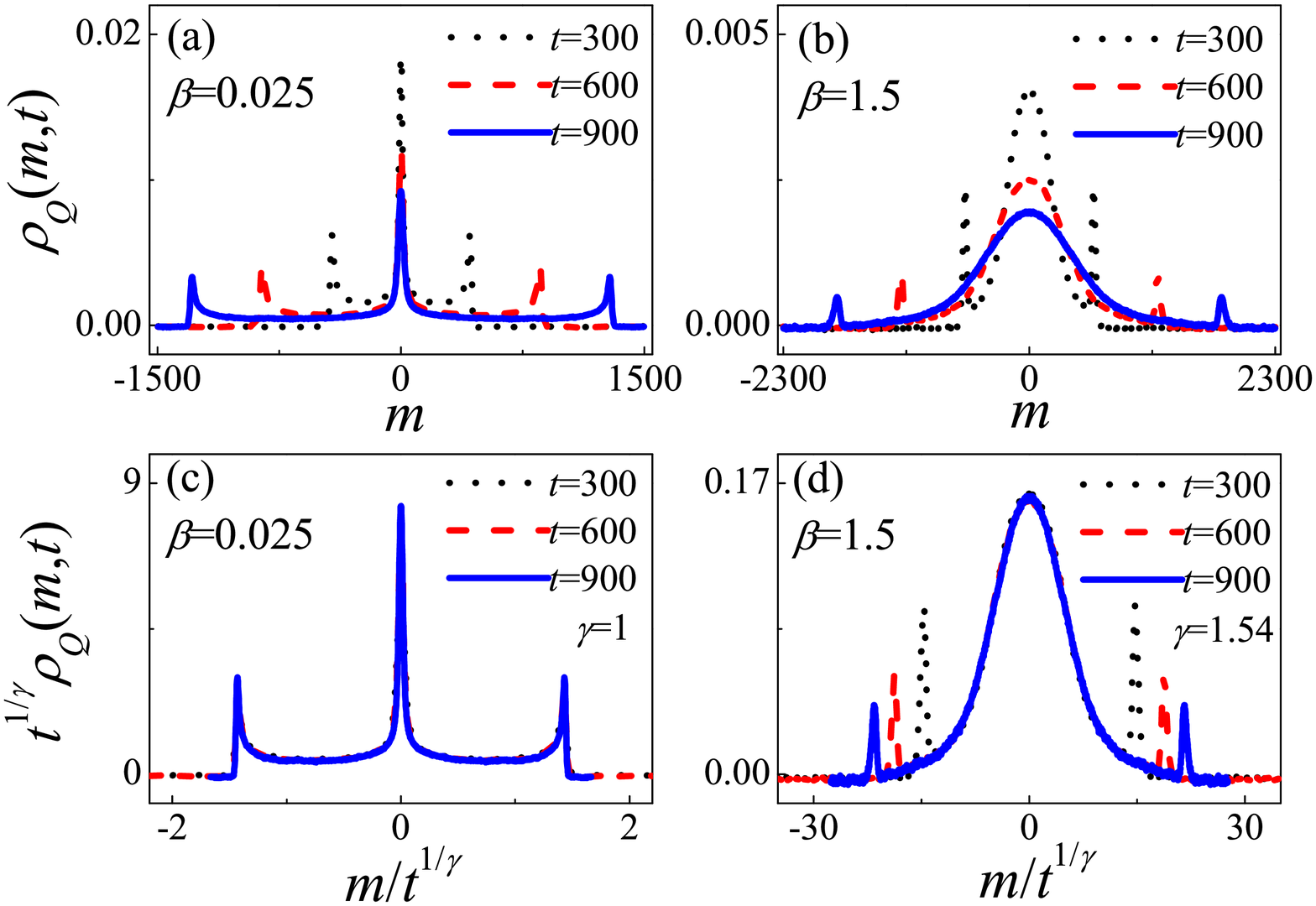}
\vskip-0.4cm \caption{\label{Fig2} (a) and (b) Profiles of $\rho_Q(m,t)$; (c) and (d) rescaled $\rho_Q(m,t)$ under formula~\eqref{scaling}. Here, three long times $t=300$ (dotted), $t=600$ (dashed) and $t=900$ (solid) and two $\beta$ values $\beta=0.025$ and $\beta=1.5$ are considered for comparison.}
\end{figure}
%%%%%%%%%%%%%%%%%%%%%%%%%%%%%%%%%%%%%%%%%%%%%%%%%%%%%%%%%%%%%%%%%%%%%%%%%%%%%%%%fig2

With such theoretical understanding, now let us turn to the
simulation results. Figure 2 presents the profiles of $\rho_Q(m,t)$
and their scaling properties for three typical times and two $\beta$
values, among which, $\beta=0.025$ denotes the system close to the
linear case; while $\beta=1.5$ we employ to represent the highly
nonlinear case. As expected, on one hand, at the relatively small
$\beta$ value, the profile of $\rho_Q(m,t)$ is mainly dependent on
the phonon dispersion relation~\eqref{dispersion} determined by
formula~\eqref{PRWdensity} and following the ballistic scaling
($\gamma=1$) [see Fig.~\ref{Fig2}(a) and (c)] if compared to the
prediction of~\cite{PRW}, especially that we see a highly localized
peak on the origin ($m=0$). Such localized peak of $\rho_Q(m,t)$
mainly stems from the special phonon dispersion relation
[Eq.~\eqref{dispersion} and Fig.~\ref{Fig1}], where as we have
already mentioned that, there is a wider $q$ domain with zero group
velocities near the Brillouin zone boundary. On the other hand, the
situation for the cases of relatively large nonlinearity is quite
different: The highly localized central peak now disappears,
instead, a common quasi superdiffusive ($\gamma=1.54$) L\'evy walks
profile can be observed [see Fig.~\ref{Fig2}(b) and (d)]. This seems
to indicate that the nonlinearity can cause the delocalization of
$\rho_Q(m,t)$ in certain domain induced by the phonon dispersion
relation. Therefore, it would be interesting to study in more detail
of such delocalization process, which may help us to fully
understand both roles of phonon dispersion and nonlinearity.
\subsection{Multiscaling in the intermediate range of nonlinearity}
%%%%%%%%%%%%%%%%%%%%%%%%%%%%%%%%%%%%%%%%%%%%%%%%%%%%%%%%%%%%%%%%%%%%%%%%%%%%%%%%fig3
\begin{figure}
\vskip-.2cm \hskip-0.1cm
\includegraphics[width=8.8cm]{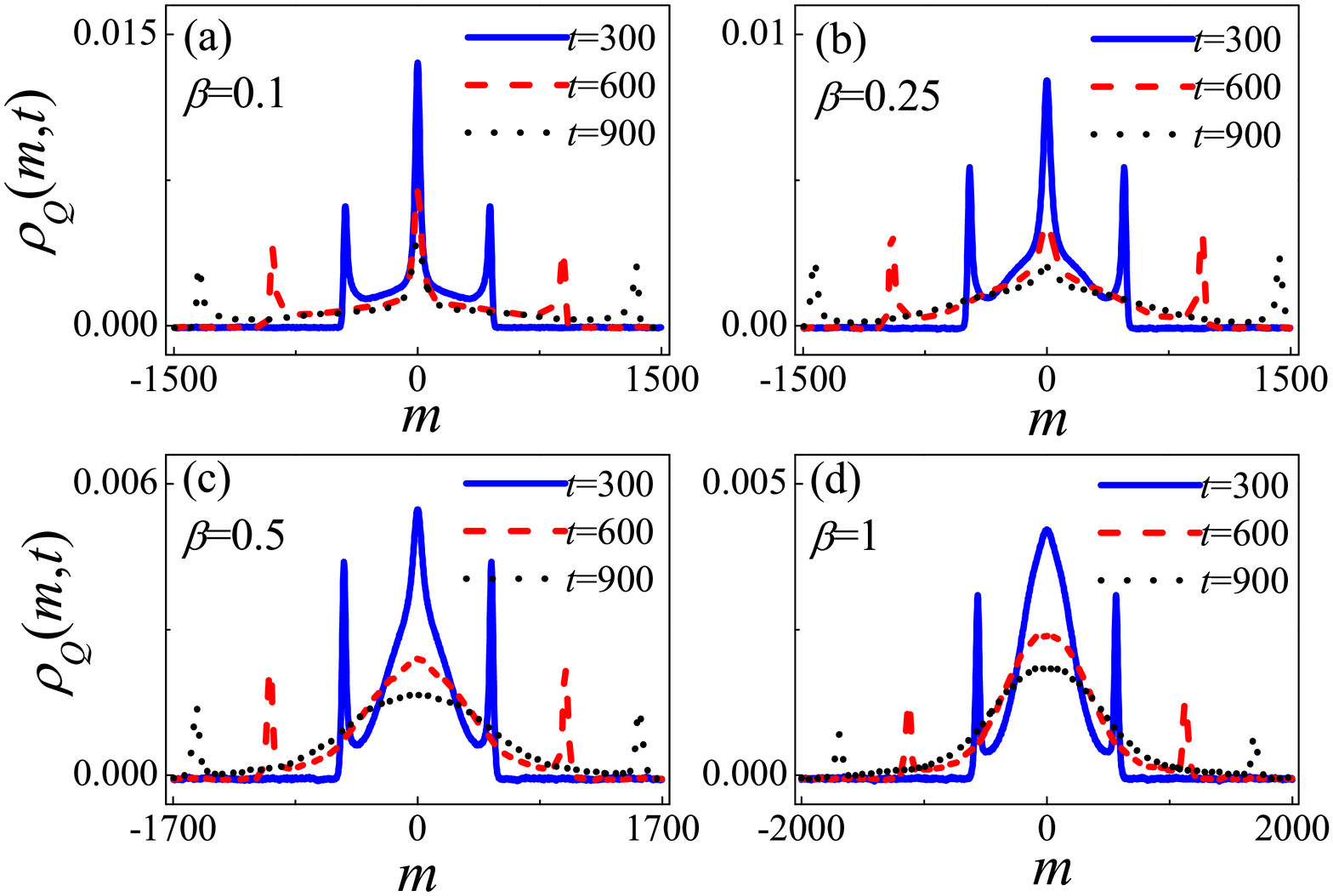}
\vskip-0.4cm \caption{\label{Fig3} Profiles of $\rho_Q(m,t)$ for
three long times $t=300$ (solid), $t=600$ (dashed) and $t=900$
(dotted) and four $\beta$ values in the intermediate range: (a)
$\beta=0.1$; (b) $\beta=0.25$; (c) $\beta=0.5$ and (d) $\beta=1$.}
\end{figure}
%%%%%%%%%%%%%%%%%%%%%%%%%%%%%%%%%%%%%%%%%%%%%%%%%%%%%%%%%%%%%%%%%%%%%%%%%%%%%%%%fig3
%%%%%%%%%%%%%%%%%%%%%%%%%%%%%%%%%%%%%%%%%%%%%%%%%%%%%%%%%%%%%%%%%%%%%%%%%%%%%%%%fig4
\begin{figure}
\vskip-.2cm \hskip-0.4cm
\includegraphics[width=8.8cm]{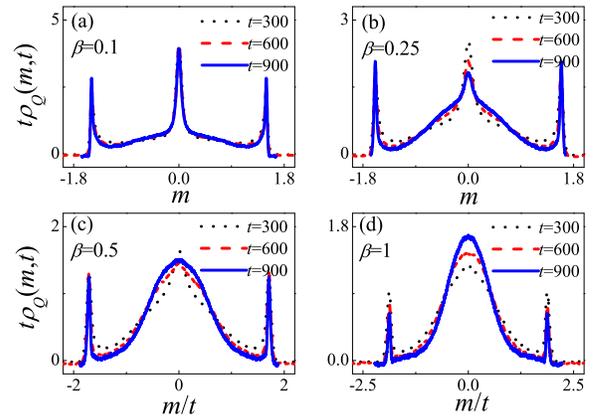}
\vskip-0.4cm \caption{\label{Fig4} Rescaled $\rho_Q(m,t)$ according
to formula~\eqref{scaling} with $\gamma=1$ for three long times
$t=300$ (dotted), $t=600$ (dashed) and $t=900$ (solid) and four
$\beta$ values in the intermediate range: (a) $\beta=0.1$; (b)
$\beta=0.25$; (c) $\beta=0.5$ and (d) $\beta=1$.}
\end{figure}
%%%%%%%%%%%%%%%%%%%%%%%%%%%%%%%%%%%%%%%%%%%%%%%%%%%%%%%%%%%%%%%%%%%%%%%%%%%%%%%%fig4
Before going on, let us first see some typical profiles of
$\rho_Q(m,t)$ in the intermediate range of $\beta$ values to gain a
preliminary impression. Figure~\ref{Fig3} presents such a result,
from which a detailed crossover from localization to delocalization
of the central peak of $\rho_Q(m,t)$ can be clearly identified.
Since such an unusual localized shape is exhibited in the center,
one can expect that the scaling formula~\eqref{scaling} is now no
longer valid, but whatever one can first use that scaling formula
with $\gamma=1$ to rescale the profiles to see how the ballistic
transport (mainly induced by the phonon dispersion dispersion) can
be destroyed by increasing the nonlinearity. For such purpose, in
Fig.~\ref{Fig4} we plot the result of rescaled $\rho_Q(m,t)$ under
ballistic scaling ($\gamma=1$). As can be seen, with the increase of
$\beta$, indeed, the distortion of ballistic scaling first
originates from the central part and then walks towards the
direction of front parts, while it should be noted that if one only
looks at the front peaks, the ballistic scaling seems still
available. Viewing this fact, it is reasonable to conjecture that
$\rho_Q(m,t)$ should at least bears two-scaling behaviors:
$\gamma=1$ for the front peaks and $\gamma \neq 1$ for others. To
further verify this multiscaling property, we here
follow~\cite{Bilinear,Bilinear-1,Bilinear-2} to study the $s$
($s>0$) order momentum of $\rho_Q(m,t)$, i.e., $\langle |m(t)|^s
\rangle = \int_{-\infty}^{\infty} |m(t)|^s \rho_Q (m,t) \rm{d}
\mit{m}$, from which for the strong anomalous diffusion process,
$\langle |m(t)|^s \rangle$ has been conjectured to follow $\langle
|m(t)|^s \rangle \sim t^{s \nu(s)}$ with $\nu(s)$ to be a
$s$-dependent exponent~\cite{Bilinear-2}; while for the ballistic
(normal) transport, it is easy to find that $\nu(s)$ will be close
to $1$ ($1/2$). Therefore, this $s$-dependent scaling exponent $s
\nu(s)$ is useful to characterize anomalous superdiffusive thermal
transport distinct from both ballistic and normal cases.
%%%%%%%%%%%%%%%%%%%%%%%%%%%%%%%%%%%%%%%%%%%%%%%%%%%%%%%%%%%%%%%%%%%%%%%%%%%%%%%%fig5
\begin{figure}
\vskip-.2cm \hskip-0.4cm
\includegraphics[width=9.8cm]{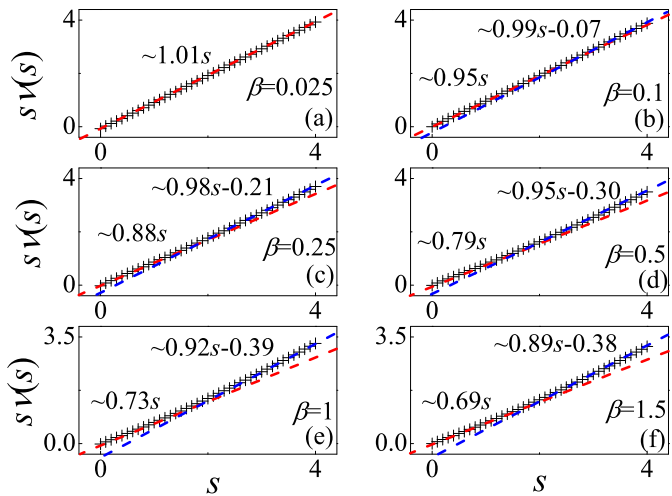}
\vskip-0.4cm \caption{\label{Fig5} $s \nu(s)$ vs $s$ for indicating the multiscaling property of $\rho_Q(m,t)$ for (a) $\beta=0.025$; (b) $\beta=0.1$; (c) $\beta=0.25$; (d) $\beta=0.5$; (e) $\beta=1$ and (f) $\beta=1.5$, respectively.}
\end{figure}
%%%%%%%%%%%%%%%%%%%%%%%%%%%%%%%%%%%%%%%%%%%%%%%%%%%%%%%%%%%%%%%%%%%%%%%%%%%%%%%%fig5

Figure~\ref{Fig5} depicts the results of $ s \nu(s)$ versus $s$
calculated from $\rho_Q(m,t)$ for several typical $\beta$ values,
from small to large. As can be seen, with the increase of $\beta$,
$s \nu(s)$ first shows a single linear scaling with $s$ [see
Fig.~\ref{Fig5}(a)], indicating $\nu(s) \approx 1$ and suggesting
the ballistic heat transport; then at about $\beta=0.1$ this single
scaling behavior is destroyed for low order $s$, eventually, a
bilinear scaling behavior starts to appear [see Fig.~\ref{Fig5}(b)];
finally, for the relatively large $\beta$ values, we see a scaling
exponent for the low order $s$, denoted by $\nu^{L}(s)$, close to
$\nu^{L}(s) \approx 0.6$-$0.7$ [see Fig.~\ref{Fig5}(f)].
%%%%%%%%%%%%%%%%%%%%%%%%%%%%%%%%%%%%%%%%%%%%%%%%%%%%%%%%%%%%%%%%%%%%%%%%%%%%%%%%fig6
\begin{figure}
\vskip-.2cm \hskip-0.4cm
\includegraphics[width=8.8cm]{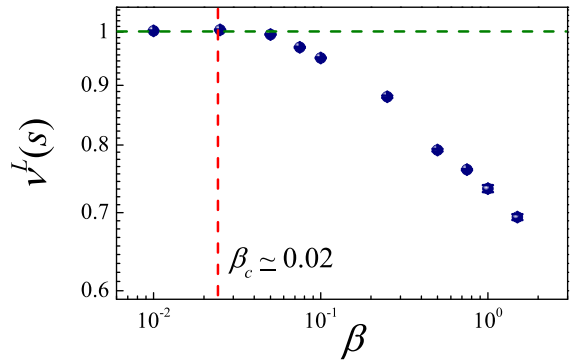}
\vskip-0.4cm \caption{\label{Fig6} The fitting values of $\nu^{L}(s)$ vs $\beta$, where the horizontal (vertical) dashed line represents $\nu(s)=1$ ($\beta_c \simeq 0.02$).}
\end{figure}
%%%%%%%%%%%%%%%%%%%%%%%%%%%%%%%%%%%%%%%%%%%%%%%%%%%%%%%%%%%%%%%%%%%%%%%%%%%%%%%%fig6

It would be worthwhile to note that such particular bilinear scaling
behavior has recently been theoretically addressed in the L\'evy
walks model within the superdiffusive regime [its density follows
the scaling formula~\eqref{scaling} with $1<\gamma<2$], where
$\gamma$ is explained as the power law exponent from the waiting
time distribution $\phi (\tau) \sim \tau^{-1-\gamma}$ of the model
(see~\cite{Bilinear,Bilinear-1} for details). In that L\'evy walks
model, $\nu^{L}(s)$ is predicted to be $1
/\gamma$~\cite{Bilinear,Bilinear-1}. Therefore, essentially
$\nu^{L}(s)$ involves important information for describing
superdiffusive transport. For this reason, we further examine the
result of $\nu^{L}(s)$ versus $\beta$ in Fig.~\ref{Fig6}. As can be
seen, a detailed crossover from ballistic ($\nu^{L}(s)=1$, thus
$\gamma=1$) to superdiffusive ($1/2<\nu^{L}(s)<1$, thus
$1<\gamma<2$) transport at a critical point of $\beta_c \simeq 0.02$
can be clearly identified. This critical point of $\beta_c$ is
consistent with the same value of $\beta_c$ found in the FPU-$\beta$
system with NN coupling only~\cite{Xiong-1}, where it has been
conjectured to be related to the strong stochasticity threshold of
the system~\cite{SST}.

For the highly nonlinear case ($\beta=1.5$ for example), bearing in
mind both the predictions of $\nu^{L}(s)=1/\gamma$ from the L\'evy
walks model~\cite{Bilinear,Bilinear-1} and $\gamma=3/2$ from the
nonlinear hydrodynamics
theory~\cite{HeatPerturbations-5,HeatPerturbations-6}, now we make a
comparison of our data with the predictions. From Fig.~\ref{Fig5}(f)
we know  $\nu^{L}(s)=0.69$, hence $\gamma$ is about $1.45$ according
to~\cite{Bilinear,Bilinear-1}. In view of the measurement errors,
this result is consistent with the prediction of
$\gamma=3/2$~\cite{HeatPerturbations-5,HeatPerturbations-6} and the
result of $\gamma=1.54$ from the scaling analysis as shown in
Fig.~\ref{Fig2}(d).
\subsection{Delocalization of the central peak}
\begin{figure}
\vskip-.2cm \hskip-0.4cm
\includegraphics[width=8.8cm]{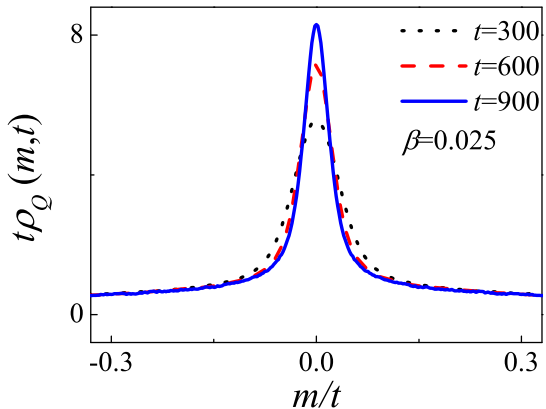}
\vskip-0.4cm \caption{\label{Fig7} Rescaled $\rho_Q(m,t)$ under the
ballistic scaling ($\gamma=1$) for $\beta=0.025$ to indicate that
for the central peak, the ballistic scaling is not perfectly valid.}
\end{figure}

After studying the whole scaling property of $\rho_Q(m,t)$, let us
consider the delocalization process of the central peak. In fact,
since a peculiar phonon dispersion relation is exhibited in the
domain near the Brillouin zone boundary (see Fig.~\ref{Fig1}), it
can be expected that the scaling behavior of this central peak is
more complicated, based on which the competition between phonon
dispersion and nonlinearity can be revealed more detailedly. To
further demonstrate this point, we first employ the result of
Fig.~\ref{Fig2}(c) once again, i.e., the rescaled $\rho_Q(m,t)$
under $\beta=0.025$, to give a more emphasis on the central peak's
scaling. As shown in Fig.~\ref{Fig7}, clearly, although the whole
shapes of $\rho_Q(m,t)$ for different $t$ can be nearly perfectly
rescaled by ballistic scaling ($\gamma=1$) [see Fig.~\ref{Fig2}(c)],
the scaling of just this central peak shows some deviations.
%%%%%%%%%%%%%%%%%%%%%%%%%%%%%%%%%%%%%%%%%%%%%%%%%%%%%%%%%%%%%%%%%%%%%%%%%%%%%%%%fig7
%%%%%%%%%%%%%%%%%%%%%%%%%%%%%%%%%%%%%%%%%%%%%%%%%%%%%%%%%%%%%%%%%%%%%%%%%%%%%%%%fig7
%%%%%%%%%%%%%%%%%%%%%%%%%%%%%%%%%%%%%%%%%%%%%%%%%%%%%%%%%%%%%%%%%%%%%%%%%%%%%%%%fig8
\begin{figure}
\vskip-.2cm \hskip-0.4cm
\includegraphics[width=9.8cm]{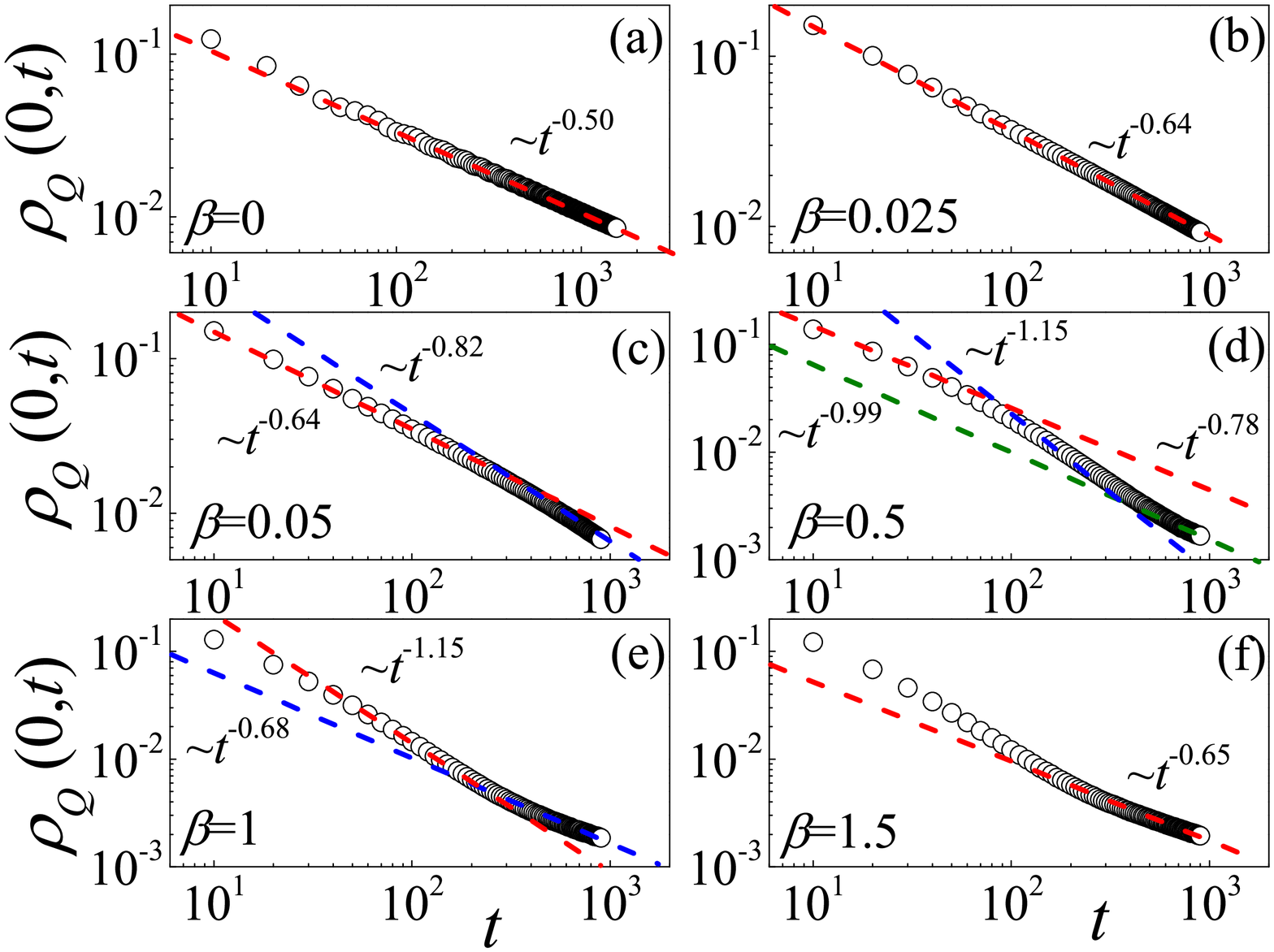}
\vskip-0.4cm \caption{\label{Fig8} $\rho_Q(0,t)$ vs $t$ for indicating the scaling property of central peak: (a) $\beta=0$; (b) $\beta=0.025$; (c) $\beta=0.05$; (d) $\beta=0.5$; (e) $\beta=1$ and (f) $\beta=1.5$, respectively.}
\end{figure}
%%%%%%%%%%%%%%%%%%%%%%%%%%%%%%%%%%%%%%%%%%%%%%%%%%%%%%%%%%%%%%%%%%%%%%%%%%%%%%%%fig8

Viewing this fact, next we investigate how the decay of this central
peak would depend on the nonlinearity. Generally, we find that
$\rho_Q(0,t)$ decays with $t$ in a power-law, i.e., $\rho_Q(0,t)
\sim t^{-\eta}$ with $\eta$ a time scaling exponent. Such relevant
results for several typical $\beta$ values are plotted in
Fig.~\ref{Fig8}, among which we note that the result of $\beta=0$ is
derived from the theoretical formula of Eq.~\eqref{PRWdensity}
suggesting the fact of $\eta=1/2$ [see Fig.~\ref{Fig8}(a)].
Interestingly, compared to the whole scaling property of
$\rho_Q(m,t)$, Fig.~\ref{Fig8} indicates that the picture of the
decay of $\rho_Q(0,t)$ for different $t$ is richer, i.e., with the
increase of $\beta$, $\rho_Q(0,t)$ first decays with one single
$\eta$ value [see Fig.~\ref{Fig8}(a)-(b)]; then at about
$\beta=0.05$, an additional $\eta$ value in the range of long $t$
emerges [see Fig.~\ref{Fig8}(c)]; after that if one increases
$\beta$ further, totally three $\eta$ values can be observed [see
Fig.~\ref{Fig8}(d)]; while eventually, for the highly nonlinear
case, the third scaling exponent in a long time, denoted by
$\eta^{L}$, seems dominated [see Fig.~\ref{Fig8}(f)]. A careful
examination of the turning point from one to two-scaling exponents
suggests a critical point of $\beta_c \simeq 0.02$, which is similar
to that demonstrated in Fig.~\ref{Fig6}. This implies that the time
scaling exponent $\eta$ of the central peak of $\rho_Q(m,t)$ also
involves key information for characterizing anomalous thermal
transport.
%%%%%%%%%%%%%%%%%%%%%%%%%%%%%%%%%%%%%%%%%%%%%%%%%%%%%%%%%%%%%%%%%%%%%%%%%%%%%%%%fig9
\begin{figure}
\vskip-.2cm \hskip-0.4cm
\includegraphics[width=8.8cm]{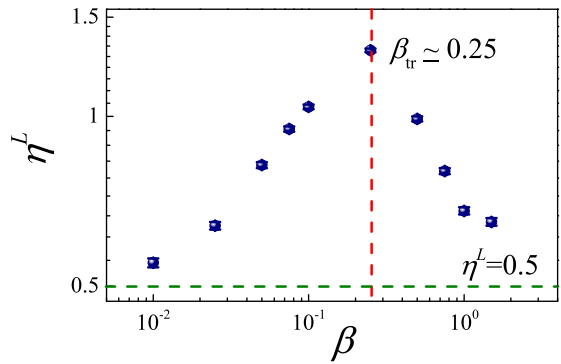}
\vskip-0.4cm \caption{\label{Fig9} $\eta^{L}$ vs $\beta$, where the horizontal (vertical) dashed line denotes $\eta^{L}=0.5$ ($\beta_{\rm{tr}} \simeq 0.25$).}
\end{figure}
%%%%%%%%%%%%%%%%%%%%%%%%%%%%%%%%%%%%%%%%%%%%%%%%%%%%%%%%%%%%%%%%%%%%%%%%%%%%%%%%fig9

Finally, as one usually concerns with the long time asymptotic
behavior of heat transport property, in Fig.~\ref{Fig9} we further
plot the result of $\eta^{L}$ versus $\beta$. One can see that, as
$\beta$ increases from $0.025$ to $1.5$, $\eta^{L}$ increases first,
then reaches its maximum value at about $\beta_{\rm{tr}}\simeq
0.25$, finally decreases down to $\eta^{L} \approx 3/2$ with the
similar value of the scaling exponent $\gamma$ shown in
Fig.~\ref{Fig2}(d). Obviously, such asymptotic decay behavior of the
central peak with $\beta$ is \emph{nonmonotonic}, which is
consistent with the same nonmonotonic variation of the
size-dependent scaling exponent of heat conductivity with
temperature in the same model~\cite{NNN}, thus further supporting
the fact that the heat conduction of this system is nonuniveral,
dependence of certain system's parameters~\cite{NNN}.
\section{Discussion}
We are particularly interested in the underlying mechanism of the
observed nonmonotonic delocalization process of $\rho_Q(m,t)$. Form
the microscopic point of view, this can be understood by the
property of intraband DBs~\cite{NNN}, which is nonmonotonically
dependent on the temperature, hence it is nonmonotonically dependent
on the nonlinearity. Eventually if the heat transport can be
understood by the picture of phonons scattered by such intraband
DBs, a nonmonotonic delocalization of $\rho_Q(m,t)$ with $\beta$
could be expected~\cite{NNN}. We here aim to explore this scattering
process of phonons from a macroscopic point of view. For this aim,
in the following we shall employ the correlation function of
momentum perturbations to reveal the information of such scattering
process.

The momentum spread described by its perturbations correlation
function $\rho_p(m,t)$ contains useful information in understanding
heat transport of momentum-conserving systems. As mentioned in Sec.
III, $\rho_p(m,t)$ may correspond to the sound modes' correlation in
hydrodynamics
theory~\cite{HeatPerturbations-4,HeatPerturbations-5,HeatPerturbations-6}.
A diffusive momentum spread has been conjectured to be the origin of
the normal heat transport observed in coupled rotator
systems~\cite{YunyunLi}. This  nonballisitic spread of momentum has
also been revealed in another special system with a double-well
interparticle potential~\cite{Xiong-2}. In the phonon random walks
theory, the momentum correlation function in linear systems has been
proved to be a quantum like wave function's real part, while this
quantum wave function itselves modulus square can represent the heat
perturbations correlation function~\cite{PRW}. A more recent
work~\cite{Cai-1} developed an effective linear stochastic structure
theory to derive the momentum spreading correlation function in the
long time limit, and demonstrated that anomalous thermal transport
is dominated by the longwavelength renormalized
waves~\cite{Cai-2,Cai-3,Cai-4}.

\begin{figure}
\vskip-.2cm \hskip-0.1cm
\includegraphics[width=8.8cm]{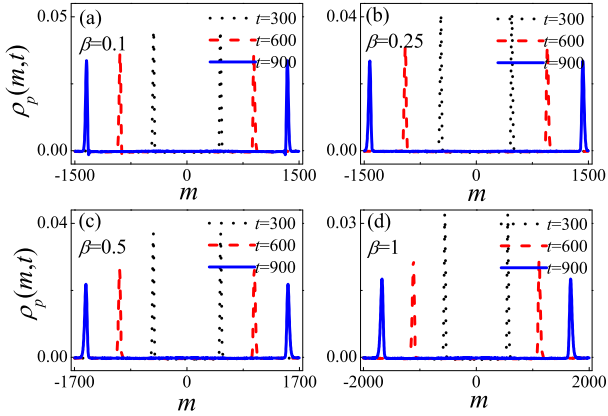}
\vskip-0.4cm \caption{\label{Fig10} Profiles of $\rho_p(m,t)$ for
three long times $t=300$ (dotted), $t=600$ (dashed) and $t=900$
(solid) and four $\beta$ values in the intermediate range: (a)
$\beta=0.1$; (b) $\beta=0.25$; (c) $\beta=0.5$ and (d)
$\beta=1$.}
\end{figure}

We begin with presenting some typical results of $\rho_p(m,t)$ in
Fig.~\ref{Fig10}. Here, three long times and four $\beta$ values the
same as those in Fig.~\ref{Fig3} are considered. As can be seen, all
the profiles of $\rho_p(m,t)$ suggest nondiffusive ballistic
behaviors, which are the evidences of anomalous heat transport. In
particular, three points can be revealed: First, the moving
velocities of the front peaks are increased with the increase of
$\beta$, which is a natural property of the systems with hard-type
anharmonicity~\cite{Xiong-3}, and can be understood from the
renormalized waves theory~\cite{Cai-2,Cai-3,Cai-4}. Second, as
$\beta$ increases, a slight broadening of the side peaks can be
identified. This broadening is related to the sound
attenuation~\cite{HeatPerturbations-4}. In the hydrodynamic
theory~\cite{ModeCoupling-1,ModeCoupling-2,Daswell-1,Daswell-2,Daswell-3,Daswell-4,HeatPerturbations-5,HeatPerturbations-6},
it has been usually suggested that there is a mode-dependent damping
coefficient $\Gamma_q \sim D |q|^ {\delta}$ in the small wavenumbers
limit, where $D$ is a damping constant and $\delta$ is a scaling
exponent. Then, the constant $D$ can be inferred from this
broadening~\cite{Cai-1}.
%%%%%%%%%%%%%%%%%%%%%%%%%%%%%%%%%%%%%%%%%%%%%%%%%%%%%%%%%%%%%%%%%%%%%%%%%%%%%%%%fig10
%%%%%%%%%%%%%%%%%%%%%%%%%%%%%%%%%%%%%%%%%%%%%%%%%%%%%%%%%%%%%%%%%%%%%%%%%%%%%%%%fig10

\begin{figure}
\vskip-.2cm \hskip-0.1cm
\includegraphics[width=8.8cm]{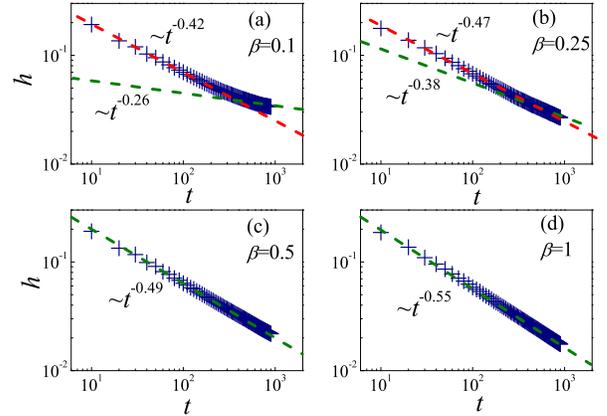}
\vskip-0.4cm \caption{\label{Fig11} The height $h$ of the front peaks of $\rho_p(m,t)$ decays with $t$: (a) $\beta=0.1$; (b) $\beta=0.25$; (c) $\beta=0.5$ and (d) $\beta=1$.}
\end{figure}

Finally, for each $\beta$ value, the front peaks of $\rho_p(m,t)$
decay with $t$ in a power law: $h \sim t^{-\delta}$, here $h$ is the
height of the peaks. This power-law exponent $\delta$ has been
conjectured to correspond to the exponent shown in
$\Gamma_q$~\cite{Cai-1}. It may also correspond to both the scaling
exponents of sound modes' correlation function predicted by the
nonlinear hydrodynamics
theory~\cite{HeatPerturbations-5,HeatPerturbations-6} and of the
heat current power spectra suggested by the mode
coupling~\cite{ModeCoupling-1,ModeCoupling-2} or mode
cascade~\cite{Daswell-1,Daswell-2,Daswell-3,Daswell-4} theory. For
this reason, we here examine detailedly the decay of this ballistic
front peak of $\rho_p(m,t)$ in some intermediate ranges of $\beta$
values (see Fig.~\ref{Fig11}). Remarkably, we find that this decay
behavior also shows a sensitive dependence of $\beta$. In
particular, around a turning point of $\beta_{\rm{tr}} \simeq 0.25$,
there is a crossover from two-scaling exponents (for small $\beta$)
to one-scaling (for large $\beta$). Eventually, for the relatively
large $\beta$ values, the values of $\delta$ seem to converge to a
value of $0.5$-$0.6$ [see Fig.~\ref{Fig11}(c) and (d)], which is
almost consistent with the value of $\delta=1/2$ predicted by the
relevant theories in the long time
limit~\cite{HeatPerturbations-5,HeatPerturbations-6,ModeCoupling-1,ModeCoupling-2,Daswell-1,Daswell-2,Daswell-3,Daswell-4}.
However, for the relatively small $\beta$ values in this
intermediate range, obviously, it seems to show some deviations.
%%%%%%%%%%%%%%%%%%%%%%%%%%%%%%%%%%%%%%%%%%%%%%%%%%%%%%%%%%%%%%%%%%%%%%%%%%%%%%%fig11
%%%%%%%%%%%%%%%%%%%%%%%%%%%%%%%%%%%%%%%%%%%%%%%%%%%%%%%%%%%%%%%%%%%%%%%%%%%%%%%%fig11

Given the same turning point of $\beta_{\rm{tr}}=0.25$, it is
reasonable to conjecture that there should be a close relationship
between the delocalization of the central peak of $\rho_Q(m,t)$ and
the decay of the side peaks of $\rho_p(m,t)$. Such a relation
revealed here is amazing, since the central peak of $\rho_Q(m,t)$ is
contributed by phonons with high wave numbers, while the side peaks
of $\rho_p(m,t)$ are represented by those with low $q$, in view of
their different group velocities. So, even that there are some
couplings between $\rho_Q(m,t)$ and $\rho_p(m,t)$, such couplings
should naturally arise at the same locations with the same $q$ (the
same group velocities)~\cite{HeatPerturbations-11}. However, our
results here seem to violate this natural intuition.

Turning back to the related theoretical models, the nonlinear
hydrodynamics theory~\cite{HeatPerturbations-5, HeatPerturbations-6}
claimed that, for sound modes, the correction from the coupling of
heat mode will vanish in the long time limit, but it decays very
slowly, so at the intermediate time scales, one could see some
physically interesting information of the heat and the sound modes'
coupling. But obviously, the nonlinear hydrodynamics theory did not
tell us that the coupling between the sound and the heat modes'
correlation can be so unusual. As to this point, we note that the
assumption of the mode cascade
theory~\cite{Daswell-1,Daswell-2,Daswell-3,Daswell-4} may present a
correct picture. The central argument the authors suggested is that,
the thermal conductivity at any (sufficiently low) frequency can be
entirely determined by the thermal conductivity and bulk viscosity
(represented by the momentum transport) at much higher frequency.
This sort of unusual coupling between the heat transport and the
momentum transport, which is the explicit basis of the mode cascade
theory~\cite{Daswell-1,Daswell-2,Daswell-3,Daswell-4}, seems to be
supported by our present results here, although the argument is
still hard to examine in more detail.
\section{Conclusion}
To summarize: We have employed the heat perturbation correlation
function $\rho_Q(m,t)$ to investigate anomalous thermal transport in
a 1D FPU-$\beta$ lattice including both the NN and NNN interactions.
After choosing an appropriate coupling ratio, we have obtained a
peculiar phonon dispersion relation which then enables us to examine
both roles of phonon dispersion and nonlinearity in more detail. It
has been found that, for relatively small and large nonlinearity,
the transport are ballistic and L\'evy walks types, respectively,
which can be well understood from the predictions of the concept of
phonon random walks~\cite{PRW} and the theory of nonlinear
fluctuating
hydrodynamics~\cite{HeatPerturbations-5,HeatPerturbations-6}. While
more interesting things take place in the intermediate range of the
nonlinearity, where we emphasize that, there both the phonon
dispersion relation and nonlinearity can play the roles. In this
intermediate range, we have found that: (i) there is a transition
from the single scaling to multiscaling for the whole profiles of
$\rho_Q(m,t)$ with a critical point of $\beta_c \simeq 0.02$; (ii)
the delocalization of the central peak of $\rho_Q(m,t)$ shows a
\emph{nonmonotonic} dependence of nonlinearity with another turning
point of $\beta_{\rm{tr}} \simeq 0.25$.

The first critical point of $\beta_c \simeq 0.02$ indicates a
crossover from ballistic to nonballistic transport and might be
related to the strong stochasticity threshold of the focused
FPU-$\beta$ systems, either or not yet including the NNN
interactions~\cite{Xiong-1}. The second turning point of
$\beta_{\rm{tr}} \simeq 0.25$ is a special feature of such systems
and seems to be related to the nonuniversal heat conduction observed
previously in the same system~\cite{NNN}. In those previous
publications~\cite{NNN}, we have conjectured that the microscopic
underlying mechanism is caused by the scattering of phonons by
intraband DBs. Here instead, we use the momentum perturbation
correlation function $\rho_p(m,t)$ to explore this phonons'
scattering process from a macroscopic point of view, which makes
this conjecture more convincing. Remarkably, we find that the time
decay behavior of the side peaks of $\rho_p(m,t)$ follows a similar
nonmonotonic $\beta$-dependent manner as those shown in
$\rho_Q(m,t)$.

Finally, we would like to point out that such a coincidence of the
properties of $\rho_Q(m,t)$ and $\rho_p(m,t)$ suggests the very
slowly decoupling process of the heat and the sound modes'
correlations claimed by nonlinear hydrodynamics
theory~\cite{HeatPerturbations-5,HeatPerturbations-6}, in particular
for the cases in the intermediate range of the nonlinearity (or at
the intermediate time scales). It also supports the assumption of
the mode cascade theory proposed by G. R. Lee-Dadswell \emph{et
al}~\cite{Daswell-1,Daswell-2,Daswell-3,Daswell-4}, i.e., the
unusual coupling between the heat transport and the momentum
transport should be taken into account. This is because the heat
mode and the sound modes are naturally considered as contributed by
the relatively high and low wave-number phonon modes with the
relatively high and low frequencies, respectively.

In short, the results presented here might provide further useful
information for our understanding of anomalous heat transport and
its coupling to the momentum transport, particularly from the
perspective of phonon dispersion relation and nonlinearity, though
understanding the detailed underlying mechanism still requires
efforts.
\begin{acknowledgments}
This work was supported by the National Natural Science Foundation of China (Grant No. 11575046); the Natural
Science Foundation of Fujian Province, China (Grant No. 2017J06002); the Training Plan Fund for Distinguished Young Researchers from Department of Education, Fujian Province, China, and the Qishan Scholar Research Fund of Fuzhou University, China.
\end{acknowledgments}

\end{document}